\documentclass[prd,tightenlines,preprint,nofootinbib]{revtex4}

\def\nc{N_{\rm c}}

\def\Tr{\,{\rm Tr}\:}

\def\hf{\frac{1}{2}}

\def\st{\begin{equation}}
\def\stp{\end{equation}}
\def\bg{\begin{eqnarray}}
\def\nd{\end{eqnarray}}
\def\nn{\nonumber}

\def\Dsl{\hbox{/\kern-.6000em\rm D}}
\def\dsl{\hbox{/\kern-.5600em$\partial$}}
\def\pxpsl{\hbox{/\kern-.5600em$p$}}
\def\ssl{\hbox{/\kern-.5600em$s$}}
\def\epssl{\hbox{/\kern-.5600em$\epsilon$}}
\def\delsl{\hbox{/\kern-.7000em$\nabla$}}
\def\lxpsl{\hbox{/\kern-.5600em$l$}}
\def\kxpsl{\hbox{/\kern-.5600em$k$}}
\def\qxpsl{\hbox{/\kern-.5600em$q$}}

\def\O{{\cal O}}

\def\x{{\bf x}}
\def\y{{\bf y}}

\def\g1{$U_Y(1)$}
\def\g2{$SU_L(2)$}
\def\g3{$SU_c(3)$}
\def\g31{$SU_c(3)\times U_{em}(1)$}
\def\g21{$SU_L(2)\times U_Y(1)$}
\def\g321{$SU_c(3)\times SU_L(2)\times U_Y(1)$}
\def\Eq#1{Eq.~(\ref{#1})}

\def\nott#1{\setbox0=\hbox{$#1$}                
   \dimen0=\wd0                                 
   \setbox1=\hbox{/} \dimen1=\wd1               
   \ifdim\dimen0>\dimen1                        
      \rlap{\hbox to \dimen0{\hfil/\hfil}}      
      #1                                        
   \else                                        
      \rlap{\hbox to \dimen1{\hfil$#1$\hfil}}   
      /                                         
   \fi}                                         %

\input epsf

\begin{document}

\title{3D ${\cal N}{=}1$ SYM Chern-Simons theory on the Lattice}

\author{Joshua W.~Elliott and Guy D.~Moore}
\affiliation{
    Physics Department,
    McGill University,
    3600 rue University,
    Montr\'{e}al, QC H3A 2T8, Canada
    }%

\date {\today}

\begin{abstract}
We present a method to implement 3-dimensional ${\cal N}{=}1$ SUSY
Yang-Mills theory (a theory with two real supercharges containing gauge
fields and an adjoint Majorana fermion) on the lattice, including a way
to implement the Chern-Simons term present in this theory.  At nonzero
Chern-Simons number our implementation suffers from a sign problem which
will make the numerical effort grow exponentially with volume.  We also
show that the theory with vanishing Chern-Simons number is anomalous;
its partition function identically vanishes.
\end{abstract}

\maketitle

\section{Introduction}
Lattice studies of supersymmetric field theories have long been
an elusive goal. The lattice regularization generally breaks the
supersymmetry and the IR limit of the lattice theory typically contains
SUSY breaking fermion and scalar mass terms, which in 4 dimensions
arise at all orders in perturbation theory.
Much effort has recently been directed toward finding ways to avoid these
terms by creative choices of lattice actions and/or generalized supersymmetry
algebras \cite{Kaplan_lattsusy,Catterall2d}.

One way around this problem is to consider minimally supersymmetric
Yang-Mills theory, which contains only gauge fields and fermions.  The
theory is then automatically (accidentally) supersymmetric provided
one can correctly implement the fermions.  Recent advances in fermion
implementations \cite{Kaplan_domainwall,Narayanan} have made it
possible to achieve this program in 4 dimensions \cite{4DSUSY}.

In this paper we instead consider 3-dimensional minimally
supersymmetric Yang-Mills theory ($N{=}1$ supersymmetry, with two real
supercharges).  This theory is free of scalars and so correct
implementation of the fermions again yields the right supersymmetric
IR limit ``accidentally.''  However, the implementation of the
fermions is quite intricate, since one must impose a Majorana
condition, and the implementation is further complicated by phases
arising both from the fermionic determinant and from a Chern-Simons
(CS) term, which is possible (and we will show, required) in this theory.

The goal of this paper is to give a recipe for studying $N{=}1$ SUSY
in three dimensions with a Chern-Simons term on the lattice.
This theory is believed to display very interesting nonperturbative
properties that make it a prime target for simulation.  In particular,
Witten has conjectured \cite{Witten3d} that the theory either preserves
or spontaneously breaks supersymmetry, depending on the value of the
Chern-Simons term.  A lattice study of the theory would then
constitute the first test (to our knowledge) of a nonperturbative
supersymmetry breaking mechanism.

The paper will be organized as follows. In section \ref{sec:continuum} we
will review the continuum action of the theory, the necessity of a
Chern-Simons term, and the anomaly condition which fixes the
Chern-Simons term to take certain half-integer values.
We will also present a proof, apparently unrecognized before, that the
theory with vanishing Chern-Simons term has a vanishing partition
function and is therefore not well defined.  
In section \ref{sec:lattice} we will describe the discretization process,
showing that the magnitude of the fermion determinant can be included
using a rooted 3-D (2 component) Wilson-Dirac fermion with SW
improvement and mass counterterm.  We then show how to extract the
Chern-Simons phase and the phase of the properly regularized rooted
determinant.
We leave the simulation itself for a
future work; the goal here is to show that such simulations can be
done and to provide the required tools.

\section{3D $N{=}1$ SYM in the Continuum}
\label{sec:continuum}
\def\prodprime{\prod_i \!' \,}

The field content of the theory consists of a gauge field and a
2-component, Majorana fermion in the
adjoint representation (gaugino).  In 4-dimensional notation, the
gaugino is a 2-component Weyl fermion which has been further reduced
from 2 complex to 2 real components by the application of a Majorana
condition (possible in 3 dimensions).  The {\em Euclidean} action is
\bg
S=\frac{1}{g^2}\int d^3x \Bigg(\frac{1}{4}F^a_{\mu\nu}F^a_{\mu\nu}+
\hf\bar{\psi}_a(\nott{D}\psi)_a\Bigg)\,.
\nd
With these conventions the eigenvalues of the Dirac operator are all
pure imaginary.
Because the fermion is in a real representation of the gauge group,
the fermionic operator possesses a doubled spectrum; if $\psi_\lambda$
is an eigenvector of $\nott{D}$ with eigenvalue $i\lambda$, then
$\epsilon \psi_\lambda^*$ (where $\epsilon=-i\sigma_2$) is also an
eigenvector with the same eigenvalue;
\bg
\sigma_\mu D_\mu \psi_\lambda & = & i \lambda \psi_\lambda \nn \\
\sigma_\mu^* D_\mu \psi^*_\lambda & = & -i \lambda \psi^*_\lambda
\nn \\
\epsilon \sigma_\mu^* D_\mu \psi^*_\lambda & = & -i \epsilon
\lambda \psi^*_\lambda \nn \\
\sigma_\mu D_\mu (\epsilon \psi^*_\lambda) & = & i\lambda
 (\epsilon \psi^*_\lambda) \,.
\nd
The Majorana condition consists of taking only one of these degenerate
sets of eigenvalues to define the fermion contribution to the path integral.
That is, in the path integral replace
\bg
\det \nott{D}=\prod_i \lambda_i
\quad\to\quad
\sqrt{\det \nott{D}}=\prodprime\lambda_i
\nd
where $\prodprime$ is defined
by taking only one eigenvalue from each degenerate pair.

We can add a Chern-Simons term to this action
in 3D provided we include an appropriate mass term for the fermions so
that SUSY is retained:
\bg
\label{eq:CS}
S_{CS}&=& -\frac{ik}{16\pi}\int_x \epsilon_{\mu\nu\rho}
\bigg(F^a_{\mu\nu}A^a_\rho-\frac{1}{3}f_{abc}A^a_\mu A^b_\nu A^c_\rho\bigg)
+ \hf \frac{k}{4\pi}\int_x\bar{\psi}_a\psi_a\\
&\equiv&-2\pi i kN_{cs}+\hf \frac{m}{g^2} \int_x\bar{\psi}\psi\,,
\qquad
m = \frac{g^2 k}{4\pi} \,.
\nn
\nd
Here $k$ is the level of the CS theory or CS coupling.
It is straightforward to check that the action is indeed invariant
under the SUSY transformations
\bg
&&\delta A^a_\mu=\bar{\alpha}\sigma_\mu\psi_a\,,\quad
\delta F^a_{\mu\nu}
=\bar{\alpha}(\sigma_\nu\partial_\mu-\sigma_\mu\partial_\nu)\psi_a\nn\\
\mbox{and}\;\;&&
\delta \psi_a=-\frac{i}{2}\epsilon_{\mu\nu\rho}F^a_{\mu\nu}\sigma_\rho\alpha\,,
\nd
with $\alpha$ the Grassman valued Majorana spinor parameterizing the SUSY
transformation. Our convention for Majorana spinors is
$\bar{s}=s^\top\epsilon$.

It has long been known \cite{Redlich:1983dv,Witten3d} that
- for gauge group SU($\nc$) -
$k$ must equal $\nc /2$ modulo an integer to avoid a gauge anomaly so that,
in particular, the theory with odd $\nc$ is ill defined for vanishing CS term.
In Sec.\ \ref{sec:anomaly} we review this argument and present a
similarly motivated argument for the theory with
even $\nc$ that implies that these
theories are also ill defined for vanishing $k$.

\subsection{Anomalies in 3D SU($ \nc $) SYM}
\label{sec:anomaly}

In \cite{Wittensu2:1982} Witten gave the first example of a theory with
a rooted determinant that is sick with a global gauge anomaly, namely 4D
SU(2) gauge theory with an odd number of left-handed fermion doublets.
The problem with this theory is that it is impossible to define the
fermionic determinant over the space of gauge connections (gauge fields
modulo all gauge transformations) such that it is both continuous and
single-valued.

We will review the 3D analog, formulated by Redlich
\cite{Redlich:1983dv} for gauge group SU($\nc$),
which is slightly simpler and is of
immediate importance for the current discussion.  The general issue is
that there are phase ambiguities in performing Grassman integrations.
This is a problem in writing a path integral unless the phase ambiguity
can be reduced to a single gauge-field independent phase, which factors
out from the partition function and cancels in determining any
correlation function.
Therefore we pick some gauge field $A_0$,
(chosen so that the Dirac operator has no zero eigenvalues) and call
its contribution to the path integral real and positive.
Then we determine the sign for any other configuration $A$ by
insisting on continuity of $\det \nott{D}$ along a path from $A_0$ to $A$.
This is possible for any gauge field because our group manifold is path
connected.

\begin{figure}[ht]
\begin{center}
$\begin{array}{c@{\hspace{.6in}}c}
\epsfxsize=2.5in\epsffile{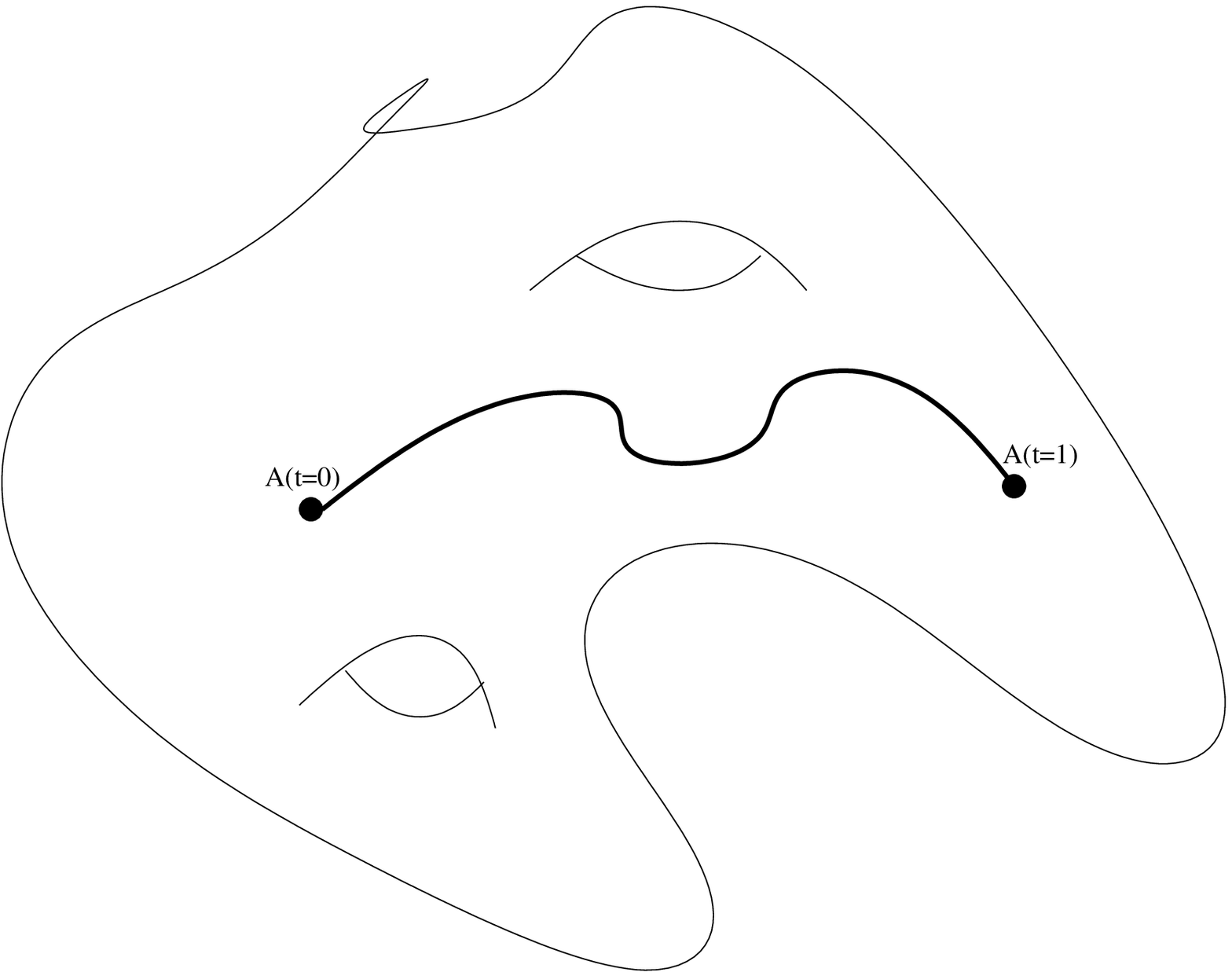}
&
\epsfxsize=1.8in\epsffile{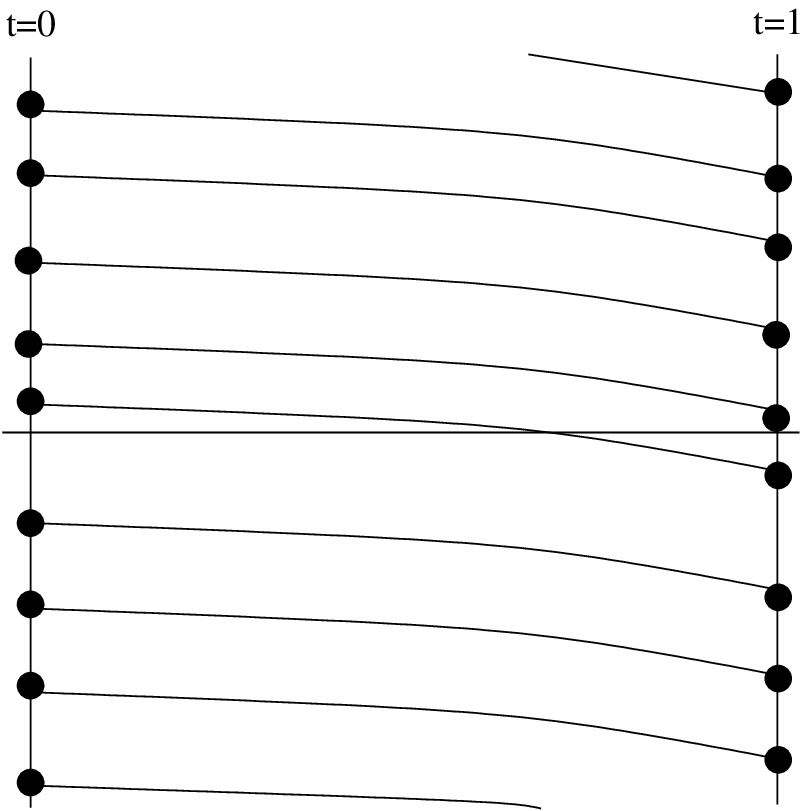}
\end{array}$
\end{center}
\caption{A path in configuration space and its associated eigenvalue
flow\label{fig:specflow}}
\end{figure}

The procedure is sketched in Fig.\ \ref{fig:specflow}:
we watch the low lying eigenvalues of the
spectrum as the path parameter $t$ is varied
from $0$ to $1$ and count the number of eigenvalue pairs that change sign
from positive to negative.  This gives the relative sign of the two
determinants in the path integral.

This prescription is unique unless the number of eigenvalue
zero-crossings depends on the path.  This can happen because the space
of connections is multiply connected; if two paths from $A_0$ to $A$
form a noncontractible loop, there is no guarantee that they lead to the
same sign choice for $\det \nott{D}(A)$.

Call the space of 3D gauge connections $X$.  It contains noncontractible
loops if the third homotopy group $\pi^3$ of the gauge group is
nontrivial.  (It is the third homotopy group because we are in 3
dimensions.)  This is the case for all SU($\nc$), for which $\pi^3={\cal
Z}$.  Consider a nontrivial loop from $A_0$ back to $A_0$.
\begin{figure}[ht]
\centerline{\epsfxsize=2in\epsfbox{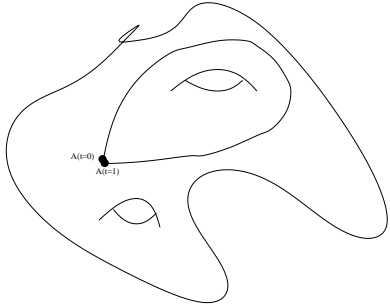}}
\caption{A closed loop in configuration space with a non-trivial winding
number \label{fig:closedloop}}
\end{figure}
This path is effectively a 4D gauge field configuration, where the space
is $S_1 \times$ 3D space and the 4D gauge fields are $A_M =
(A^\mu,A_4=0)$.  The path is noncontractible if the instanton number of
this gauge field configuration is nonzero.  The four dimensional Weyl
determinant has a number of zero modes determined by the Atiyah-Singer
theorem \cite{Atiyah-Singer}; for the fundamental representation this
is 1 and for the adjoint representation this is $2\nc$.
The 4D zeros correspond to zero crossings, and therefore sign flips, of
the 3D fermionic determinant.  If the number of sign flips in traversing
the loop is odd, then the definition of the determinant cannot be both
continuous, nontrivial, and single valued.

For our case this is relevant because we want the square root of the
Weyl determinant; the $2\nc$ zero crossings become $\nc$ zero crossings
when we choose one from each pair of eigenvalues, and this leads to a
sign flip if $\nc$ is odd.

There is an additional sign if the theory is defined with a nonzero
Chern-Simons term.
The CS term picks up a factor of $2\pi\nu$ in traversing a path of
instanton number $\nu$, so the path integral picks up an overall factor of
\bg
(-1)^{\nu\nc}\exp(i2\pi \nu k )\,.
\nd
This implies that, in order to avoid a gauge anomaly, $k{=}\nc/2$ modulo
an integer.  Only certain (half-integer) values of the Chern-Simons term
are allowed, and in particular, the theory with $\nc$ odd and
vanishing CS term is ill defined.

We will now show that the supersymmetric theory with $\nc$ even and
Chern-Simons coefficient $k=0$ (and therefore fermion mass of zero) is
also anomalous, a point which to our knowledge has
not been noticed before.
Consider a configuration ${A(x)}$ and its parity dual ${A'(x)=-A(-x)}$.
We claim that these give canceling contributions to
the partition function if the Chern-Simons term is absent and the
fermion mass is zero.
They clearly have the same bosonic action,
so we must show only that their rooted fermion determinants are
equal and opposite.  We define the sign of the rooted determinant for
configuration ${A(x)}$ to be positive and
connect them with a path from ${A(x)}$
to the trivial vacuum and
from the trivial vacuum to ${A'(x)}$
via the parity dual of this path (see again Fig.\ \ref{fig:specflow}).

\begin{figure}[ht]
\centerline{\epsfxsize=2in\epsfbox{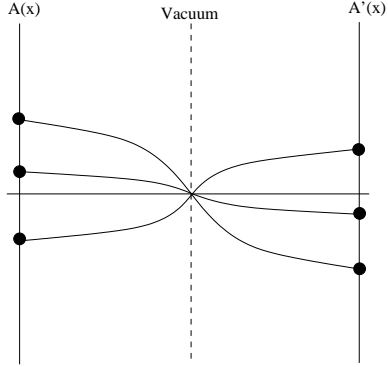}}
\caption{Flow of almost zero mode eigenvalues of the Dirac operator as
the configuration is varied between parity conjugate gauge fields on a
path through the vacuum. \label{fig:su2flow}}
\end{figure}

The Dirac operator for the trivial vacuum configuration has
$(\nc^2-1)$ pairs of zero eigenvalues (we implicitly work on a
torus with standard boundary conditions). There are then $(\nc^2-1)$ pairs
of eigenvalues that cross zero at the vacuum configuration. Furthermore,
if $n_+$ pairs cross zero somewhere on the path {\em between} ${A(x)}$ and the
vacuum, than the number of pairs $n_-$ that cross zero between the vacuum
and ${A'(x)}$ will be the same.
The total number of
pairs which change sign in going from ${A(x)}$ to ${A'(x)}$ is therefore
$(\nc^2-1) + 2 n_+$. This is odd, so
the fermion rooted determinant flips sign and the configurations
give canceling contributions to the partition function, which vanishes
identically.
An example of the eigenvalue flow for gauge group SU($2$) is shown in
Fig.\ \ref{fig:su2flow} for the case $m{=}0$.

This problem is avoided at nonzero Chern-Simons term
because ${A(x)}$ and ${A'(x)}$ have opposite Chern-Simons number and so
enter the partition function with opposite phase, rather than
canceling.  Similarly, at nonzero fermion mass the eigenvalues of the
Dirac operator are complex and introduce nonzero phases which are
opposite between the configuration and its parity dual.

\subsection{Regularization dependence of determinant}
\label{subsec:returnloop}

\begin{figure}
\centerline{\epsfxsize=3.5in\epsffile{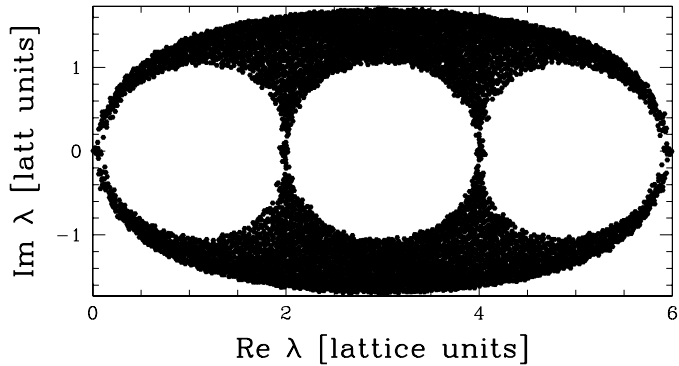} }
\caption{\label{fig:return}
Eigenvalues for 20 Wilson fermion configurations in a $8^3$ box at
$g^2 a=0.5$, illustrating the ``return loops'' in the complex plane.}
\end{figure}

It is necessary to clarify what is meant by the ``continuum'' fermionic
determinant.  The issue is that continuum theories are always defined as
limits of discrete (or otherwise regularized) theories and in a discrete
theory with volume
regularization there are a finite number of eigenvalues for the Dirac
equation.  In traversing a closed loop with nonzero instanton number, we
just saw that a nonzero number of eigenvalues cross from negative to
positive value.  Since the final configuration is the same as the
starting one, it has the same spectrum.  Since there are a finite number
of negative eigenvalues, there must be some compensating flow of
positive eigenvalues to negative, somewhere in the complex plane.  In
other words, eigenvalues must ``return'' somewhere in the
complex plane.  We illustrate this for Wilson fermions in
Fig.~\ref{fig:return}.  The figure superimposes the eigenvalue spectra,
in the complex plane, of 20 quenched gauge field configurations in an
$8^3$ box with lattice spacing $g^2 a=0.5$.  Each dot is a pair
of eigenvalues (the pairing of the spectrum discussed in the last
section occurs for both the Wilson and overlap lattice implementations
of the Dirac operator).  The
spectrum of eigenvalues for $\nott{D}$ parallels the imaginary axis near
zero but bends out into the complex plane for large eigenvalues $\lambda
\sim 1/a$ and forms a loop, so eigenvalues moving from negative to
positive values ``push'' eigenvalues around the loop to reappear at
negative values.  These extra vanishing-imaginary-part eigenvalues are
the lattice fermion doublers which have been pushed out into the complex plane
by the Wilson term; for the Wilson action in 3 dimensions there are
actually 3 extra places where eigenvalues cross zero imaginary part,
corresponding to $(\pi,0,0)$, $(\pi,\pi,0)$, and $(\pi,\pi\,\pi)$ type
doublers.

The problem is that this ``return loop'' will contribute to the
partition function even in the continuum limit.  Because it represents
deeply UV physics, its contribution to the partition function
must be representable in an effective IR description in terms
of local effective operators.  The 3D theory has only one marginal
operator, which is the Chern-Simons term.  Therefore the return loop can
(and will) induce a Chern-Simons term, but will not otherwise change the
infrared description (in the small $a$ limit).  It is easy to see that the size
of this Chern-Simons term is fixed by the rate of spectral flow near
zero; the number of eigenvalues which circle around the loop must be the
number needed to refill the negative eigenvalues when the spectrum flows
upwards.  The sign of this extra Chern-Simons term depends on whether the
return loop is at positive or negative imaginary part, which is a
regularization detail.  The desired continuum theory is the one in which
this regularization effect has been removed by a Chern-Simons number
counterterm.

\section{The Discretization}
\label{sec:lattice}

This section describes the discretization of $N{=}1$ SYM with a
Chern-Simons term in 3D.  The theory contains nontrivial phases; that
is, different gauge field configurations contribute to the partition
function with different complex phase as well as different magnitude.
The plan is to treat this using the Edinburgh method; one studies the
theory on the lattice by building a Markov chain sample weighted by the
magnitude of the action $|\exp(-S) \sqrt{\det \nott{D}}|$ 
and then includes the phase as part of the
observable.  Phase cancellation reduces the statistical power in a
volume dependent way.  Therefore our implementation is on the same
footing as finite chemical potential simulations in QCD; they work in
principle, but whether they work in practice depends on how severe the
phase cancellation problem turns out to be.

The implementation consists of two parts; the real part of the bosonic
action and magnitude of the fermionic determinant, and the phase.  The
real bosonic action is completely standard.  We describe the magnitude
of the determinant first, then the Chern-Simons part of the phase, then
the phase in the determinant.

\subsection{Fermion Implementation}
\label{sec:fermionimplement}

Though there may be some advantages to using an overlap fermion
implementation to do the simulation, we believe that the numerical
simplicity of the Wilson implementation makes it a much more sensible
choice.  The usual problems with Wilson fermions are less severe than in
4 dimensions; chiral symmetry is a non-issue because there is no such
thing in 3 dimensions, and the additive renormalization of the mass is
well behaved because the theory is super-renormalizable.  One can easily
work at a fine lattice spacing where the problem of exceptional
configurations is well under control (note that we are only interested
in the theory at finite fermion mass, since as we just argued the
massless theory is anomalous).

\begin{figure}[ht]
\centerline{
\begin{picture}(350,130)
  \thicklines
  \put(0,60){\line(1,0){130}}
  \put(70,0){\line(0,1){120}}
  \put(20,65){\line(1,0){45}}
  \put(20,65){\line(0,1){45}}
  \put(20,110){\line(1,0){45}}
  \put(65,110){\vector(0,-1){40}}
  \put(120,65){\vector(-1,0){40}}
  \put(75,65){\line(0,1){45}}
  \put(75,110){\line(1,0){45}}
  \put(120,65){\line(0,1){45}}
  \put(20,10){\line(1,0){45}}
  \put(20,10){\line(0,1){45}}
  \put(20,55){\vector(1,0){40}}
  \put(65,10){\line(0,1){45}}
  \put(75,10){\line(1,0){45}}
  \put(75,10){\vector(0,1){40}}
  \put(75,55){\line(1,0){45}}
  \put(120,10){\line(0,1){45}}
        \put(135,55){$\mu$}
        \put(65,125){$\nu$}
        \put(5,30){$U_\nu$}
        \put(5,85){$U_\nu$}
        \put(40,115){$U_\mu$}
        \put(87,115){$U_\mu$}
        \put(124,85){$U^\dag_\nu$}
        \put(90,0){$U^\dag_\mu$}
        \put(124,30){$U^\dag_\nu$}
        \put(40,-2){$U^\dag_\mu$}
        \put(145,58){$\displaystyle \;=\; \frac{1}{4}\Tr\bigg(
U^\dag_\mu (x)U^\dag_\nu (x{+}\mu)U_\mu(x{+}\nu) U_\nu(x)+\,\dots \bigg)$}
        \end{picture}}
\caption{Clover field strength for the Sheikholeslami-Wohlert improvement
term.\label{fig:clover}}
\end{figure}
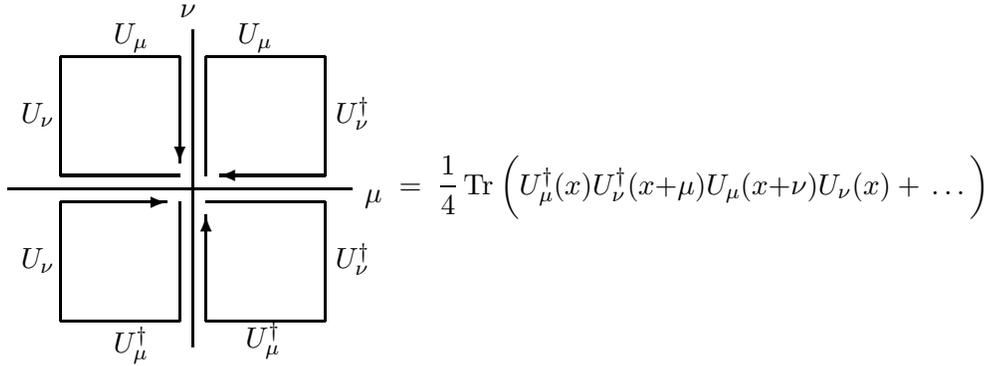

The fermionic action reads
\bg
S_{W}&=& a^3\sum_x\bar{\psi}_x \left(m_0+\frac{3r}{a}\right)\psi_x\nn\\
&&-a^3 \sum_{x,\,\mu}\bar{\psi}_x
\frac{(r-\sigma_\mu)U_\mu(x)\psi_x
+(r+\sigma_\mu)U^\dag_\mu(x{-}\mu)\psi_{x{-}\mu}}{2a} \,.
\nd
We can improve the convergence of the spectrum to the continuum limit
from $O(a)$ to $O(a^2)$ via the 3D analog of the
Sheikholeslami-Wohlert term \cite{Sheikholeslami:1985ij} (SW) term
\bg
S_{sw}= a^3\sum_{x,\, \mu, \,\nu}
\frac{r c_{sw}}{16}\bar{\psi}_x [\sigma_\mu,\sigma_\nu]
\bigg(P_{\mu\nu}(x)-P^\dag_{\mu\nu}(x)\bigg)\psi_x
\nd
Here $P_{\mu\nu}(x)$ is the {\em average} of the 4 plaquettes in the $\mu\nu$
plane as shown in Fig.\ \ref{fig:clover}.
This improvement is probably necessary to implement simulations at
reasonable lattice spacings.

Since the 3D theory is superrenormalizable we
need only the tree level determination of the SW
coefficient, $c_{sw}{=}1$, to remove $O(a)$ corrections other than mass
renormalizations.  Much of the complication of the improvement in 4D is
thus avoided in the 3D version.

\begin{figure}[ht]
\centerline{\hfill
\epsfxsize=2.8in\epsffile{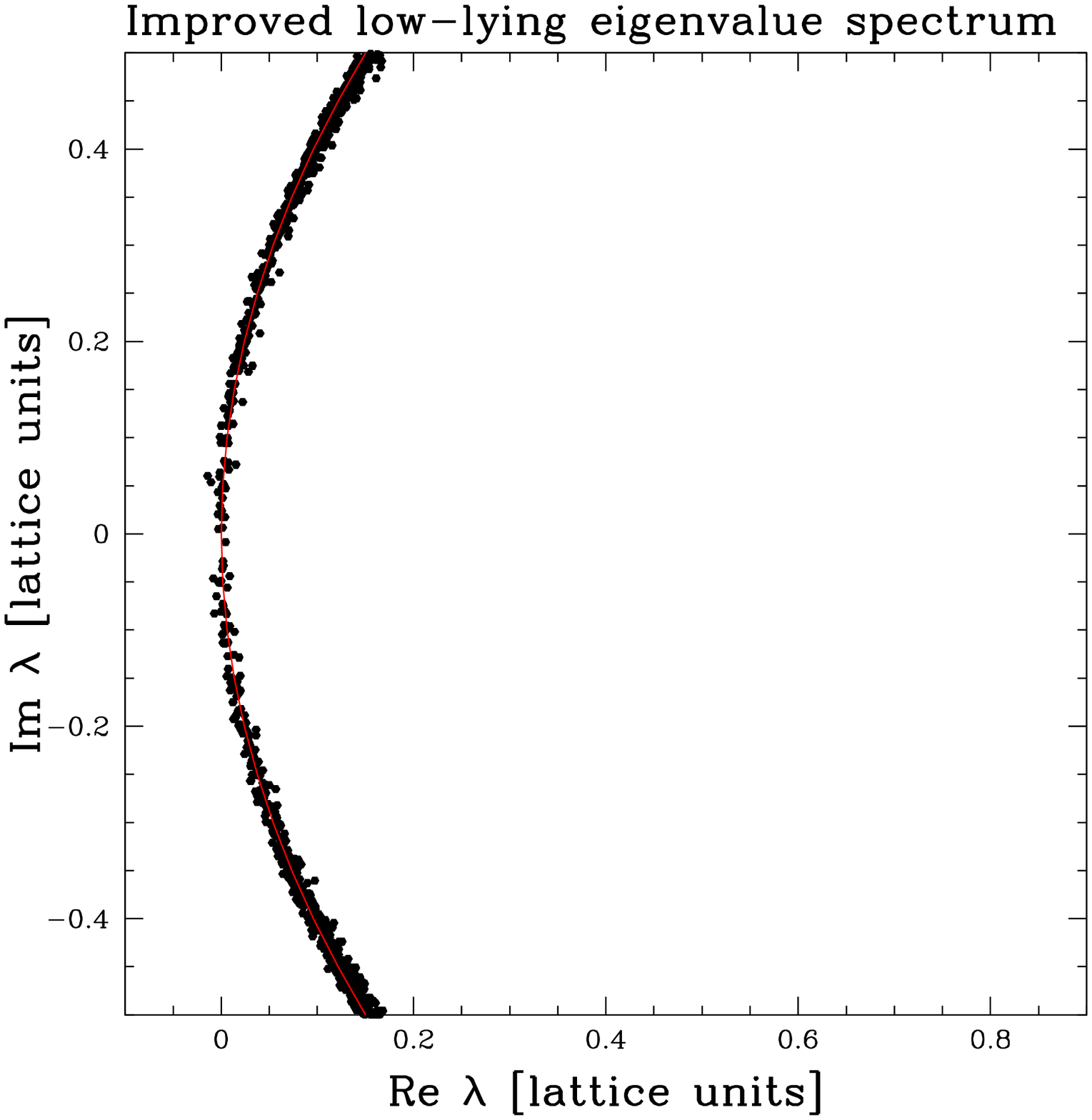}
\hfill \hfill
\epsfxsize=2.8in\epsffile{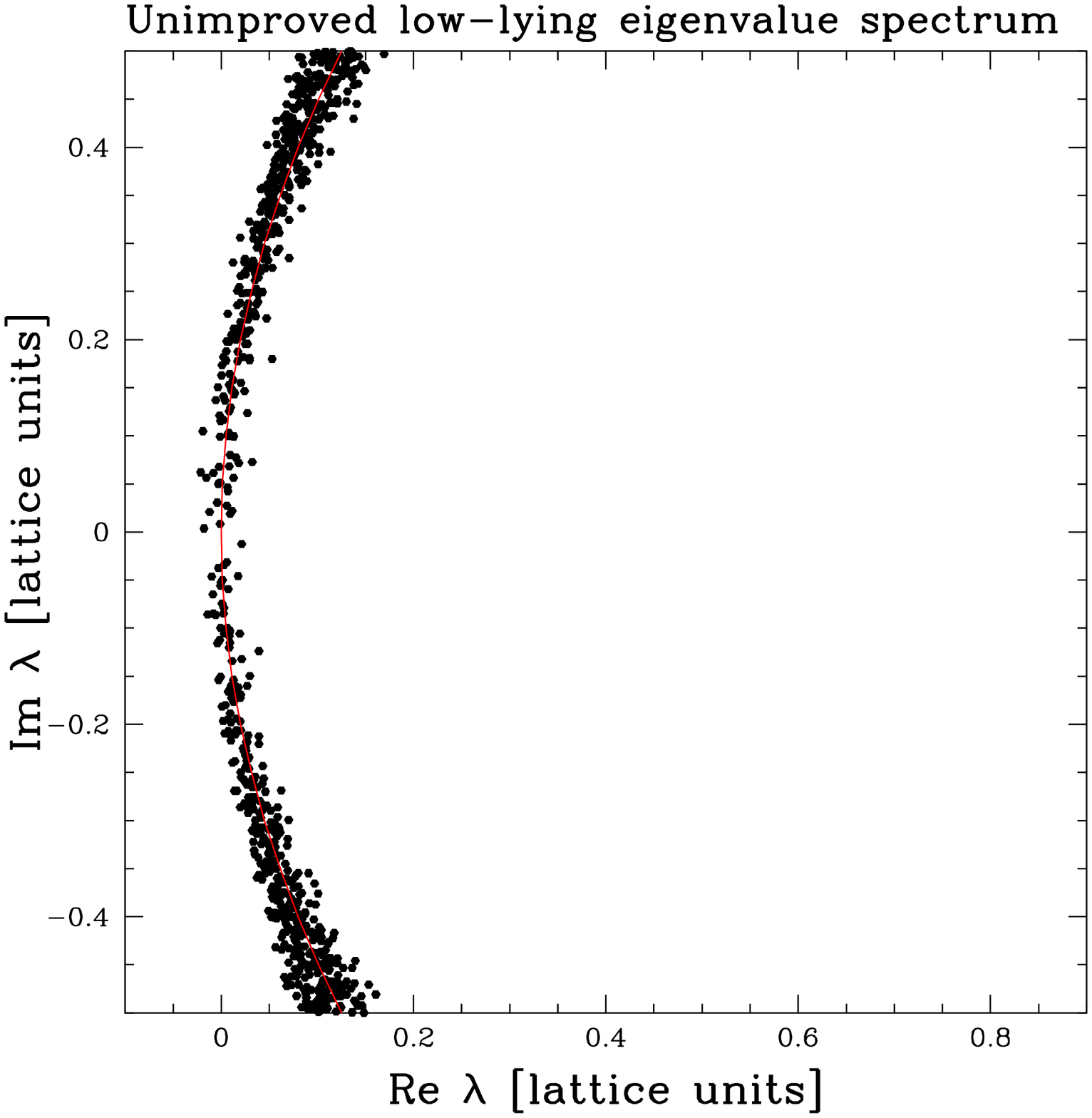}
\hfill}
\caption{\label{fig:improvement}Physical branch of the spectrum of the 
Wilson-Dirac operator; 20 superimposed configurations (left) with and
(right) without improvement ($14^3$ lattices with $g^2 a = 0.5$)}
\end{figure}

The improvement to the spectral
properties is quite dramatic, though it parallels 
closely the improvement
in 4D so we refer the interested reader to, for example,
\cite{Gattringer:1998ab}
for analysis of 4D Dirac operator spectra with
improved and unimproved actions in a situation which is fairly analogous
to the cases we consider in 3D. Fig.\ \ref{fig:improvement} shows an 
example of the improvement to the physical branch of the spectrum 
for 20 configurations on a $14^3$ lattice with $g^2 a = 0.5$. Improvement 
``squeezes'' the eigenvalues toward the solid line--a parabola
incorporating the ``bending'' in the complex plane present in the tree
level Dirac operator due to the $k$ dependence of the Wilson term.
The spectra shown are 
shifted appropriately so that both represent a fermion determinant for the 
case of zero physical mass. 

The disadvantage of Wilson fermions is that we must tune the fermion
mass to remove an additive correction.  To consistently improve the
theory to eliminate all $O(a)$ errors we need to do this at the two loop
level (see \cite{elliottmoore} for a discussion).  Here we present the
calculation of the one-loop mass counterterm; in practice both one and
two loop counterterms can be determined quite easily numerically by
analyzing the low-lying eigenvalues in the Dirac operator spectrum at a
few different lattice spacings.

The SW term modifies only the $\bar\psi A \psi$ vertex.
Following
closely the 4D treatment of \cite{Capitani:2002mp} we determine
the new three point vertex to be (with $\bar{\psi}(p)$, $\psi(p')$ and
$A(k{=}p'{-}p)$)
\bg
\Big(V_\mu\Big)^a_{bc}(p,p')=-g T^a_{bc}\Bigg(
i\sigma_\mu\cos\frac{(p{+}p')_\mu}{2}+\frac{r}{2}\widetilde{(p{+}p')}_\mu
+ \frac{c_{sw}r}{4}[\nott{\hat{k}},\sigma_\mu]
\cos\frac{k_\mu}{2}
\Bigg)\,.
\nd
The one-loop bare mass counterterm required to tune the theory to its
SUSY limit, for Wilson coefficient $r=1$, is
\bg
\delta m=\frac{g^2 C_A}{4\pi}(-2.3260)=\frac{C^2_A}{2\pi\beta}(-2.3260)\,,
\nd
with $C_A$ the first Casimir of the adjoint representation of the gauge
group - $C_A{=}\nc$ for SU($\nc$).

To eliminate all $O(a)$ errors it is also necessary to compute a 1-loop
fermionic wave-function renormalization (which can be interpreted as an
$O(a)$ multiplicative renormalization of the mass) and a 1-loop
renormalization of the field strength (or equivalently of $g^2$, or
equivalently of the lattice spacing).  The contribution of bosonic loops
to the gauge coupling renormalization was computed in
\cite{Oa2}, and the fermionic contribution
is not difficult. In the same notation as that previous work, 
\bg
{\cal L}_{\rm latt}(x)=\frac{2}{Z_gg^2a^4}\sum_{i<j}\Tr(1-P_{ij}(x))
+Z_\psi\bar{\psi}\left(D_W+D_{sw}+\frac{Z_m}{Z_\psi}m 
+\frac{\delta m}{Z_\psi}\right)\psi\,,
\nd
where it was determined that 
\bg
Z^{\rm bosonic}_g= 1- \frac{ag^2}{4\pi}\left(\frac{2\pi}{9\nc}(2\nc^2-3)
+\nc \left(\frac{37\xi}{12}-\frac{\pi}{9}\right)\right)
\quad\mbox{with}\quad\xi=0.152859325\,,
\nd
we have 
\bg
\frac{Z_m}{Z_\psi} & = & 1-\frac{ag^2\nc}{4\pi}e_m\qquad\mbox{with}
\quad e_m=2.75066732(8)\\
\mbox{and}\qquad
Z^{\rm total}_g & =& Z^{\rm bosonic}_g 
   -\frac{ag^2\nc}{4\pi}e_g\qquad\mbox{with}
\quad e_g=0.204254488(7)\,.
\nd
To treat $N_{\rm f}$ fundamental representation fermions rather than one
adjoint with a Majorana condition, replace $\nc \rightarrow C_{\rm f}$
in the expression for $Z_m/Z_\phi$ and change
$\nc \rightarrow N_{\rm f}$ in the expression for $Z_g$ (the $1/2$ trace
normalization group theory factor cancels a factor of 2 from not
imposing the Majorana condition).

As already mentioned, $\delta m$ is easily measured nonperturbatively 
by - for example - 
looking at the lowest lying eigenvalues of the Dirac operator for a range of 
$\beta$ values and fitting to find what value of $\delta m$ is needed to
get the zero imaginary part eigenvalues to have vanishing real part as
well. To tune out the O($a$) error analytically would require
a 2-loop determination of the fermion self energy, which is much more 
difficult. 

The numerical implementation of the magnitude of the determinant is
achieved by including $(\det {\cal D}^\dagger {\cal D})^{1/4}$ in the
path integral, which can be accomplished by conventional pseudofermion
techniques.  There should be no issues with locality (we believe)
because the operator's spectrum is doubled, so the fourth root is
actually well defined (up to a phase).

\def\NCS{N_{\rm CS}}
\subsection{Evaluation of Chern-Simons number}
\label{sec:CSterm}

Here we detail how to determine the Chern-Simons number of a
3-dimensional lattice configuration.  Our approach is borrowed from an
investigation of the electroweak sphaleron rate by one of us
\cite{broken_sphaleron}, but we summarize it here for completeness.

The literal definition of Chern-Simons number $\NCS$ in terms of the
integral of $F\wedge A - \frac{1}{3} A \wedge A \wedge A$, given in
\Eq{eq:CS}, is too gauge
dependent and lattice spacing sensitive to be of much use, so we use
instead the following (gauge invariant) properties of Chern-Simons
number, which can be taken as a definition of $\NCS$, modulo an integer:
\begin{itemize}
\item
$\NCS$ for the vacuum is 0 (modulo an integer); and
\item
the $\NCS$ difference between two configurations is (modulo an integer)
equal to the integral of $\epsilon^{\mu\nu\alpha\beta} \Tr F_{\mu \nu}
F_{\alpha \beta} / 32\pi^2$ along a
path through configuration space connecting those configurations.
\end{itemize}
The second point requires some clarification.
As previously discussed, given two 3D configurations $A^i_{1}(\x)$,
$A^i_{2}(\x)$, one can find a path connecting them through gauge field
configuration space; that is, one can find
$A^i(\x,\tau)$ with $\tau \in [0,1]$ an affine parameter and
$A^i(\x,0)=A^i_{1}(\x)$, $A^i(\x,1)= A^i_{2}(\x)$.  We may think of
the path as a 4-dimensional gauge field configuration, with $\tau$ as
the fourth coordinate, $D_0 \equiv D_{\tau}$, and $F_{0i} = i
[D_0,D_i]$.  The $\NCS$ difference between two configurations is
the integral
\begin{equation}
\NCS(A_2) - \NCS(A_1) =
\int d^3 x \int_0^1 d\tau \epsilon_{ijk} \frac{ \Tr F_{0i} F_{jk}}
     {8\pi^2} \,.
\end{equation}
We have not specified how $A_0$ is to be chosen along
the $\tau$ direction, but this turns out not to matter; a different
choice leads to a change in $F_{0i}$ of $D_i \delta A_0$, but the $D_i$
can be integrated by parts onto $F_{jk}$ and $\epsilon_{ijk} D_i F_{jk}$
vanishes by the Bianchi identity.  The idea is then to define $\NCS$ for
a configuration (modulo an integer) by finding a path from that
configuration to the vacuum and integrating $F\tilde{F}$ along that
path.

The complication in using this approach to define $\NCS$ on the lattice
is that there is no lattice definition of the field strength
$F_{\mu\nu}$ which satisfies the Bianchi identity; therefore the
procedure is ambiguous.  Further, no specific prescription for choosing
$A_0$ generically leads to an integer $\NCS$ around a closed loop.  There are
mathematically rigorous \cite{Luscher82,Woit85,Stone84} and numerically
implementable \cite{Hetrick,DeGrand} methods to find the integer value
around closed loops, but these are not helpful here, since we really
want the value on a path with distinct beginning and end points.  The
trick instead
is to note that the problems with lattice implementations of
$F \tilde{F}$ arise when the fields are ``coarse'' (plaquettes far from
identity, most of excitations at the lattice spacing scale).  We can
define $\NCS$ uniquely by choosing any unique prescription for a path
from a configuration to the vacuum.  We can ensure that the result
is as close as
possible to the continuum meaning of $\NCS$ if our unique prescription
is one which quickly smooths out the lattice-spacing scale fluctuations
in the gauge field configuration.  The early ``smoothing out'' part of
the path then contributes a small UV ``lattice artifact''
contribution to $\NCS$ and the remaining path gives
a contribution which closely resembles the
continuum value of $\NCS$ for this configuration.

A good
choice is the ``cooling path'' or gradient decent under the energy,
\begin{equation}
D_\tau A^i(x,\tau) = - \frac{\partial  H[A(\x,\tau)]}
   {\partial A^i(x,\tau)} \, , \qquad
H[A] \equiv \int d^3 y \; \frac{1}{2} \Tr F^2_{ij}[A(\y,\tau)] \, .
\end{equation}
We use a Symanzik \cite{Symanzik} or ``rectangle-improved'' definition
of $H$ and of the field strength appearing in $F\tilde F$, as described
in \cite{broken_sphaleron}, which also shows extensive tests of the
approach.

To summarize, the procedure is to determine the Chern-Simons number of a
lattice configuration by integrating $F \tilde{F}$ along the ``cooling''
or gradient-decent path through configurations to the vacuum.  The
procedure is unique and gives an answer which is continuous over the
space of gauge field configurations except at a ``sphaleron''
separatrix, where it is discontinuous by (almost exactly) an integer.
If Markov-chain configurations are tightly enough sampled, one can
determine the integer part by continuity.

On fine lattices this definition of Chern-Simons number should correctly
reproduce the continuum notion up to corrections suppressed by two
powers of the lattice spacing.

A straightforward alternative method to implement integer Chern-Simons
numbers would be to evaluate the phase in the determinant of a
fundamental representation Wilson
fermion with a negative mass of order the lattice spacing (so the origin
of the complex plane is inside the leftmost ``circle'' in
Fig.~\ref{fig:return}).  This method would be numerically much less
efficient, since the numerical effort in taking a determinant rises as
the third power of the number of lattice points or $a^9$, while the
approach presented only grows worse as $a^5$ and can be reduced to $a^3$
through the careful use of blocking \cite{broken_sphaleron}.

\subsection{Fermionic phase}

Since in practice only the {\em magnitude}
of the rooted fermion determinant can
be included dynamically in a lattice simulation,
we must still describe a prescription for assigning to the
Wilson-Dirac operator
a phase for each field configuration in order to complete the
prescription for the discretization of the theory.

One approach to do this would be to explicitly evaluate the (complex)
determinant of ${\cal D}$ for each configuration in the Markov chain and
then halve the phase, fixing the sign ambiguity by using a tightly
sampled Markov chain and demanding continuity.  Then one must subtract
the Chern-Simons phase contribution from the ``return loop'' discussed
in Subsection \ref{subsec:returnloop}.  This approach is correct but
numerically expensive.

After correcting for the Chern-Simons term induced by the ``return
loop'' this phase is dominated by the contributions of the low lying
eigenvalues, so we can develop a more efficient procedure by focusing on
determining the phase arising from these eigenvalues.
Since this point is key to our approach we should explain it in a little
detail.  Look again at Fig.~\ref{fig:improvement}.  In the continuum
limit the eigenvalues will lie, not on an arc, but on a straight line
with real part $m$ and imaginary part set by the eigenvalue under the
$\nott{D}$ operator (for the free theory, by $k$).  At large eigenvalue
the weak-coupling approximation is valid (since the theory is
super-renormalizable this is true by a power of the eigenvalue).  The
density of eigenvalues therefore scales with the free theory density of
states, $k^2 dk$.  An eigenvalue's phase difference from $\pm
\frac{\pi}{2}$ is $\tan^{-1}(m/k)$, so naively the phase arising from
large eigenvalues could be large.  What is important, though, is the
phase difference configuration by configuration, and this becomes small,
essentially because the large eigenvalues do not change very much as a
function of the gauge field.  To see this,
note that the large eigenvalues represent short-range physics.  The
influence on the effective IR behavior can be expanded as Wilsonian
renormalization of effective IR operators.  The lowest order
parity-odd operator is the Chern-Simons term; all others are higher by
at least two powers of derivatives and therefore suppressed
by at least $g^4/k^2$.  Since we are allowing $O(a^2)$ errors, the only
operator we need to incorporate correctly from the large $k$ eigenvalues
is therefore their contribution to the Chern-Simons term.  Therefore our
strategy will be to include low eigenvalues' phases explicitly and to
determine the phase contribution of large eigenvalues in terms of their
contribution to an effective Chern-Simons term.

\begin{figure}[ht]
\begin{center}
$\begin{array}{c@{\hspace{.6in}}c}
\epsfxsize=1.2in\epsffile{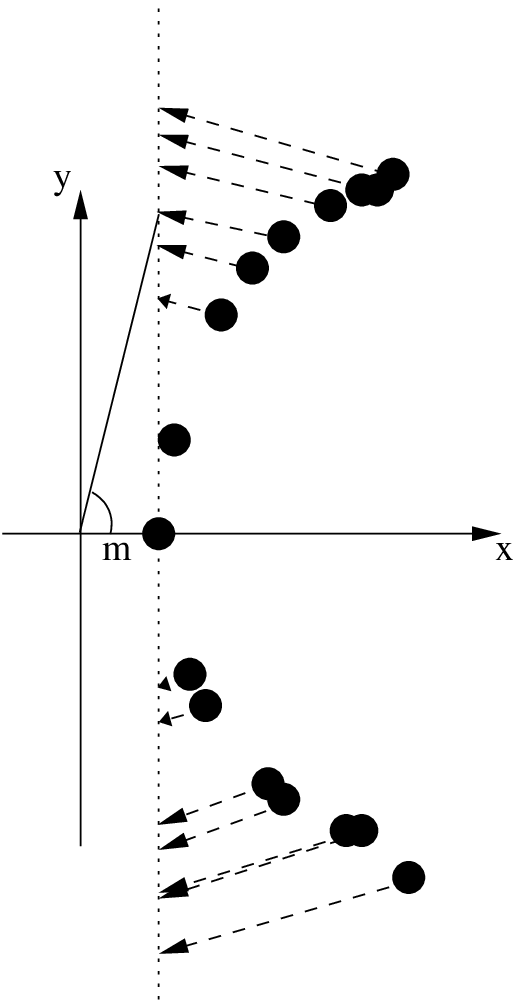}
&
\epsfxsize=2.2in\epsffile{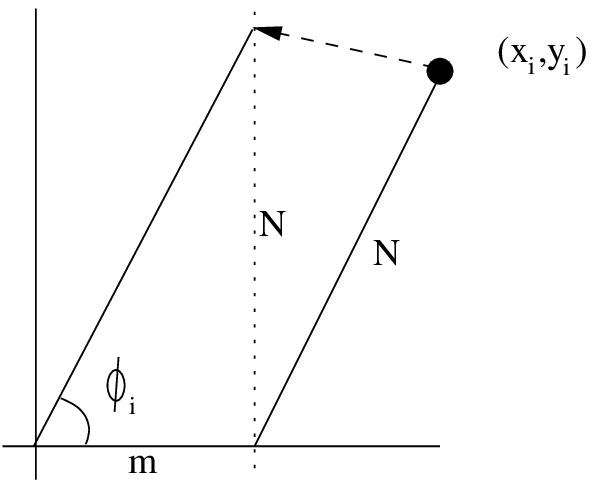}
\end{array}$
\end{center}
\caption{Sketch of the projection for the phase prescription.
\label{fig:phase_project}}
\end{figure}

We can extract the smallest $M$ eigenvalues using the Arnoldi method at
much less numerical effort than is required to determine the full
determinant.  Each such eigenvalue $\lambda_i$
of the Wilson-Dirac operator with renormalized mass $m$
takes a value in the complex
plane, $(r_i,i_i)$.  In the continuum limit the real parts $r_i$
always equal $m$.  At finite lattice spacing there will be $O(g^4 a^2)$
and $O(i^2 a^2)$ deviations in the real part.  (There would be
$O(g^2 a)$ deviations arising from the dimension 5 operator $\bar\psi
\sigma_{\mu\nu} F^{\mu\nu} \psi$ had we not included the SW term.)  We
``clean'' the low lying eigenvalues by projecting them to the $r=m$
axis, see Fig.~\ref{fig:phase_project}.  Each eigenvalue then
contributes a phase
\bg
\phi_i=\tan^{-1}\left(\frac{\sqrt{(r_i-m)^2+i^2_i}}{m}\right)\,.
\nd
This amounts to projecting the physical branch of the Wilson-Dirac spectrum
onto the axis of the continuum
spectrum and then calculating the phase as sketched in
Fig.\ \ref{fig:phase_project}.
The idea is to incorporate the phase of all eigenvalues for which the
angle $\phi_i$ lies in a range $[-\phi_{\rm max},\phi_{\rm max}]$. 
In the $a\rightarrow 0$ limit we must take $\phi_{\rm max} \rightarrow
\frac{\pi}{2}$.  This requires more eigenvalues at finer lattice
spacing; this can be made more efficient by using the shifted Arnoldi
method.

The Markov evolution of the gauge fields $U_\mu$ will move around the
eigenvalues so that eigenvalues regularly move in and out of the
``window'' in which we explicitly include them.  When an eigenvalue goes
above or below $\phi_{\rm max}$, the phase we determine will abruptly
change by $\mp \phi_{\rm max}$.
Therefore each configuration along the Markov chain must be reasonably
close to the last, so that these phases can be determined by continuity.
The change in phase between neighboring configurations $U_{1,2}$  
contributed by all eigenvalues lying above $\phi_{\rm max}$ 
{\em and} below $-\phi_{\rm max}$ is simply an
effective Chern-Simons term, as discussed above.  The size of the
contribution can be determined by considering the amount of spectral
flow due to a changing Chern-Simons number, and is well approximated by
$2(\frac{\pi}{2}-\phi_{\rm max})\nc \,(\NCS(U_2)-\NCS(U_1))$.
We have confirmed this in quenched simulations, for instance by
analyzing the $\phi_{\rm max}$ dependence of the procedure and seeing
that this contribution ensures independence on this artificial
parameter. If we choose the vacuum 
with, for example, $\NCS=0$ to have zero phase, then this prescription 
uniquely determines a phase for all configurations. 

\subsection{The sign (or phase) problem}

Now that the appropriate implementations of the fermionic field content
and the Chern-Simons term
have been detailed there remains no fundamental barrier to the simulation
of the theory and so we turn our attention to an important technical issue
of the simulation.

The lattice simulation consists of replacing the path integral
by a sum over a finite set of link field configurations
that are distributed with a probability given by the Boltzmann factor for
the theory, $|\exp(-S[U])|$. For
complex action, the ensemble average for an observable $\O$ will be
\bg
\label{eq:phase}
\langle \O\rangle \approx
\frac{\sum_{i=1}^N \O(U_i) e^{i\phi_i}}{\sum_{i=1}^N e^{i\phi_i}}\,,
\nd
where $\phi = \arg \exp(-S[U])$.
Obviously this leads to cancellations between link configurations
from the sampling, and thus to a reduction in statistics.  That is, for
a sample of $N$ independent configurations, the error in the numerator
scales as $\sqrt{N}$; but the denominator will be smaller than $N$, so
the error in the operator will not be $1/\sqrt{N}$.
This problem could be eliminated by performing ``phase quenching'' on
the theory, but this does uncontrolled damage to the theory which in our
case we believe is severe.  Therefore we must face this phase
cancellation issue.

To determine how bad the phase cancellations will be,
we start by looking at the theory with very large fermion mass, so that
the effect of integrating out the fermion is well approximated by a shift to
the Chern-Simons coefficient, $k\to k-\nc/2$. In large volumes, we expect that
$N_{cs}$ will be
Gaussian distributed around zero. The degree of phase
cancellation is determined by how badly our determination of $1$ in the
denominator of \Eq{eq:phase} is
``suppressed'': all measurables must be scaled by the result of the
partition function which we evaluate as $\langle 1 \rangle$ with the
average replaced by a sum on configurations with phases.  Fluctuations
go as $1/\sqrt{N}$ with $N$ the number of configurations (unless each
configuration contributes better statistics--this depends on the
measurable).  The point is that these fluctuations are to be compared
with an average value which suffers phase cancellation.  We estimate
this by doing the (Gaussian) integral over $N_{cs}$ to find how big the
partition function actually is:
\bg
\langle 1 \rangle = \int \frac{dN_{cs}}{\sqrt{2\pi \chi}}
\exp(-i 2\pi k N_{cs}) \exp(-N_{cs}^2/2\chi)
\nd
with $\chi$ the variance of $N_{cs}$ and $k$ the effective
Chern-Simons coefficient. The integral gives
\st
\exp( - 2\pi^2 k^2 \chi)\,.
\label{eq:statistics}
\stp

\begin{figure}[th]
\centerline{\epsfxsize=3in\epsfbox{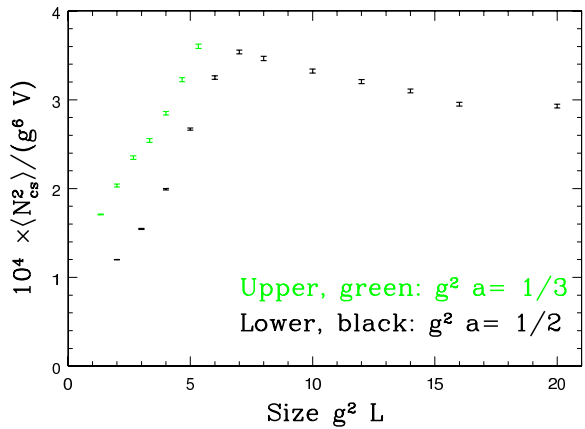}}
\caption{(color online)
  Variance of $\NCS$ as a function of volume, for two lattice
  spacings, in quenched SU(2) gauge theory and using the definition of
  $\NCS$ presented in the text. \label{fig:chidep}}
\end{figure}

Fig.\ \ref{fig:chidep} shows the
dependence of $\chi/V$ on the volume of the lattice, in quenched
simulations.
As the figure shows, going to a finer lattice induces a constant shift
in $\chi$, and the volume dependence becomes weak
for boxes larger than about $8/g^2$.  At this value, on a ``reasonably
fine'' lattice of $g^2 a=0.5$ [$\beta = 8$], the $\NCS$
variance is about .175, leading to a sign problem induced loss of
statistical power of order $\exp(-2\pi^2 k^2 \times .175)
= 1/30$ for $k=1$.  This volume is therefore at the limit of
practicality.  Statistical power falls exponentially if we try to make
the volume any larger.  However, a box of length $8/g^2$ is very
large. The theory has a mass gap and the large volume limit should be
approached rapidly in a box a few times longer than the longest
correlation length.  For the SU(2) theory, the lightest
glueball mass is $m_{0^{++}} \simeq 1.66 g^2$ \cite{Teper}
and the inverse correlation length involved in the Debye mass is
$\simeq 1.14 g^2$ \cite{Philipsen}.  These both suggest that the
dominant physics is on quite short scales $\sim 1/g^2$, though the
string tension suggests a longer correlation length $1/\sqrt{\sigma}
\simeq 3/g^2$ \cite{Teper}. Therefore this volume is probably sufficient
to effectively achieve the continuum limit.

The sign problem grows more severe at large $k$, as indicated by
\Eq{eq:statistics}.
Fortunately the mass scales with $k$, so we can reduce the volume as $V
\propto 1/k^3$ in the large $k$ limit.  We must also tighten the lattice
spacing $a$ to keep $ka$ fixed, and since $\langle \NCS^2 \rangle$ turns
out to have a linear UV divergence (which causes the lattice spacing
dependence observed in Fig.~\ref{fig:chidep}), this means that $\chi$
will scale as $V/a \sim k^{-2}$.  In the large $k$ limit the severity of
the sign problem therefore approaches a finite limit.

Fig.~\ref{fig:chidep} is based on quenched configurations.
When we include the effects of dynamical fermions, we expect the situation
at small $k$
to improve for two reasons. The first is that inclusion of the magnitude of
the determinant in the Boltzmann factor
suppresses configurations with large $\NCS$
(an effect we have observed in preliminary quenched simulations).
This is not surprising since we know that Dirac operators for
configurations with half-integer $\NCS$ (sphalerons) have zero eigenvalues.
We thus expect this suppression to be more effective at smaller fermion
masses. We will not attempt to quantify this statement further.

\begin{figure}[ht]
\centerline{\epsfxsize=3.5in\epsfbox{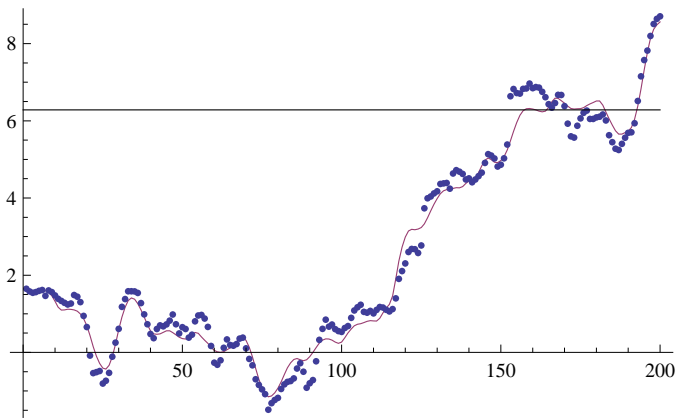}}
\caption{Evolution of the phase of the rooted fermion determinant over
a Markov chain of length 200. The continuous line
is $2\pi N_{cs}$. The system transitions through a sphaleron
and subsequently fluctuates around the
topologically distinct vacuum at $2\pi$. \label{fig:det_CS_phase}}
\end{figure}

The second reason is that the phase contributed by the determinant
partially cancels against the phase from the Chern-Simons term. For
large fermion mass, for example,
we know from continuum methods that integrating out the fermion gives a shift
in $k$ or, equivalently, a contribution to the partition function
$\exp(i\pi\nc\NCS)$.  Furthermore,
we have confirmed for the case of $\nc{=}2$ Wilson-Dirac fermions -- with the
prescription described above for assigning determinant phases -- that
this is approximately the
phase of the rooted fermion determinant for the relatively small
masses ($k{=}1,2,3$) of interest here. An example is shown in Fig.\
\ref{fig:det_CS_phase} for $k{=}1$. In this case the cancellation of the
phase in the partition function is nearly exact, and the phase problem
disappears.  Finally we remark that, in the algorithm for determining
$\NCS$, one can begin the integration of $F\tilde{F}$ after a short
amount of cooling, eliminating the contribution from the most UV modes.
This eliminates a lattice artifact ``noise'' contribution to $\chi$ but
leaves that from interesting IR physics intact. 

\section{Conclusion}
\label{sec:conclusion}

Three dimensional minimally supersymmetric gauge theory can be
implemented on the lattice.  As we showed, this theory necessarily
involves complex phases; the theory with vanishing Chern-Simons term and
massless fermions is anomalous because the path integral is odd under
parity transformations and the partition function therefore vanishes
identically.

Our implementation has concentrated on numerical efficiency.  Lattice
spacing corrections should first occur at $O(a^2)$.  Fermions need only
be implemented using the clover-improved Wilson method, not the much
more expensive overlap method.  The Chern-Simons phase can be determined
without reference to fermionic operators, and the phase in the rooted
Dirac determinant is determined by finding only the low lying
eigenvalues using the Arnoldi method, rather than by taking the full
determinant.  This efficiency is important because the theory suffers
from a sign problem which will make it difficult to take the large
volume limit.  We are guardedly optimistic that the sign problem will
not be as severe as might be feared.  First, the phase of the fermionic
determinant and of the Chern-Simons term are of opposite sign and
partially cancel.  Second, when the Chern-Simons term is large, the
theory is massive and the volume requirement should be reduced.

Now that an implementable method has been presented, it would be very
interesting to study 3D ${\cal N}{=}1$ SYM on the lattice.  In
particular it would greatly improve our insight into nonperturbative
supersymmetry breaking if we could study and (presumably) verify
Witten's conjectures regarding spontaneous SUSY breaking in this theory
\cite{Witten3d}. 

\subsection*{Acknowledgements}

We would like to thank Joel Giedt for useful conversations. JE 
thanks the University of Chicago physics 
department for their hospitality while this work was being prepared. 
This work was partially supported by
grants from the National Science and Engineering Research Council of
Canada (NSERC) and by 
le Fonds Qu{\'e}b{\'e}cois de la Recherche sur la Nature et les
Technologies (FQRNT).

\end{document}